\documentclass[conference]{IEEEtran}
\usepackage{cite}
\usepackage{amsmath,amssymb,amsfonts}
\usepackage{algorithmic}
\usepackage{graphicx}
\usepackage{textcomp}
\PassOptionsToPackage{hyphens}{url}
\usepackage{hyperref}
\usepackage{xcolor}
\usepackage{etoolbox}
\usepackage{footmisc}
\def\BibTeX{{\rm B\kern-.05em{\sc i\kern-.025em b}\kern-.08em
    T\kern-.1667em\lower.7ex\hbox{E}\kern-.125emX}}

\makeatletter
\newcommand{\linebreakand}{%
\end{@IEEEauthorhalign}
\hfill\mbox{}\par
\mbox{}\hfill\begin{@IEEEauthorhalign}
}
\makeatother

\newtoggle{draftmode}
\settoggle{draftmode}{false}

\iftoggle{draftmode}{
	\newcommand{\todo}[1]{\marginpar{\textcolor{red}{TODO\\\footnotesize #1}}}
	
	\newcommand{\bnote}[1]{\textcolor{teal}{ #1 }}
	\newcommand{\hnote}[1]{\textcolor{cyan}{ #1 }}
	\newcommand{\lnote}[1]{\textcolor{magenta}{ #1 }}
	\newcommand{\mnote}[1]{\textcolor{orange}{ #1 }}
	\newcommand{\snote}[1]{\textcolor{violet}{ #1 }}
}{ 
	\newcommand{\todo}[1]{}
	\newcommand{\bnote}[1]{}
	\newcommand{\hnote}[1]{}
	\newcommand{\lnote}[1]{}
	\newcommand{\mnote}[1]{}
	\newcommand{\snote}[1]{}
	\hypersetup{hidelinks}
}
	
\newcommand{\affiliationTUD}{Chair of Applied Cryptography \\ \textit{TU Darmstadt, Germany}\\}
\newcommand{\affiliationFS}{Frankfurt School Blockchain Center \\ \textit{Frankfurt School of Finance}\\ \textit{and Management, Germany}\\}

\begin{document}

\title{DeFi-ning DeFi: Challenges \& Pathway
}

\author{\IEEEauthorblockN{Hendrik Amler}
\IEEEauthorblockA{\affiliationTUD hendrik.amler@tu-darmstadt.de}
\and
\IEEEauthorblockN{ Lisa Eckey}
\IEEEauthorblockA{\affiliationTUD lisa.eckey@tu-darmstadt.de}
\and
\IEEEauthorblockN{Sebastian Faust}
\IEEEauthorblockA{\affiliationTUD sebastian.faust@tu-darmstadt.de}
\linebreakand
\IEEEauthorblockN{Marcel Kaiser}
\IEEEauthorblockA{\affiliationFS marcel.kaiser@fs-blockchain.de}
\and
\IEEEauthorblockN{Philipp Sandner}
\IEEEauthorblockA{\affiliationFS philipp.sandner@fs-blockchain.de}
\and
\IEEEauthorblockN{Benjamin Schlosser}
\IEEEauthorblockA{\affiliationTUD benjamin.schlosser@tu-darmstadt.de}
}

\maketitle

\begin{abstract}
The decentralized and trustless nature of cryptocurrencies and blockchain technology leads to a shift in the digital world. The possibility to execute small programs, called smart contracts, on cryptocurrencies like Ethereum opened doors to countless new applications. One particular exciting use case is decentralized finance (DeFi), which aims to revolutionize traditional financial services by founding them on a decentralized infrastructure. We show the potential of DeFi by analyzing its advantages compared to traditional finance. Additionally, we survey the state-of-the-art of DeFi products and categorize existing services. Since DeFi is still in its infancy, there are countless hurdles for mass adoption. We discuss the most prominent challenges and point out possible solutions. Finally, we analyze the economics behind DeFi products. By carefully analyzing the state-of-the-art and discussing current challenges, we give a perspective on how the DeFi space might develop in the near future.
\end{abstract}


\section{Introduction}
\label{sec:introduction}
Blockchain and distributed ledger technology (DLT) have gained huge popularity since the development of Bitcoin~\cite{nakamoto} over a decade ago. The introduction of a distributed, open and trustless ledger enables censorship-resistant borderless financial transactions between users. Beyond simple financial transactions, many DLTs support scripts for their transactions, allowing users to define complex rules and conditions for payments. Some blockchains like Ethereum~\cite{ethereum} even allow payments to depend on the execution of Turing-complete programs, so-called smart contracts. A plethora of traditionally centralized financial instruments are now being deployed and used on distributed blockchain systems using smart contracts. 
These financial tools and services are often referred to as \emph{Decentralized Finance (DeFi)}. 

The fundamental innovation of DeFi is similar to blockchains: reducing trust by replacing centralized platforms with a decentralized system. The resulting system is considered trustless, meaning that no single party must be trusted for storing funds and sending transactions. The decentralized nature also reduces the influence of individuals on the fees and conditions for using the system which is instead transparently managed by the supply and demand. Additionally, DeFi systems are open to anyone. In particular, this means individuals can also take roles like lenders, which were traditionally in the hands of large players, like banks. 

As DeFi products are built on smart contracts, multiple DeFi products can be composed by letting smart contracts interact with each other. This allows developers to build even more flexible and powerful tools. This connectivity of DeFi is often also called \emph{financial Lego}, and is viewed as an especially promising feature. Of course, composability also leads to more complexity for users and developers and has resulted in various spectacular security breaches, as we overview in this report. Notice that this is, in particular, dangerous because, unlike in centralized systems, there is no way to undo transactions in case of fraud. Once deployed, smart contracts cannot be changed easily and all funds sent to them will be processed exactly as programmed, which is not always the same as intended by the users. Countless examples of faulty smart contracts show that users risk their funds to be stolen, frozen or misused.
But despite these risks, the demand for DeFi services is unbroken, reaching new highs in fall 2020. In this work, we give an overview of some of the most pressing challenges that developers and users of DeFi systems face and discuss possible solutions. We believe that our work can be an first starting point for researchers and developers that are interested in the DeFi space. 

\subsection{Contribution}
\label{subsec:contribution}
We provide a definition for DeFi and discuss its advantages in comparison to traditional finance in Section~\ref{sec:potential}. We formulate how general properties of the distributed blockchain technology (which have been priorly only discussed in the context of single use cases and protocols, e.g., in~\cite{zile2018blockchain}, \cite{swan2017anticipating} and \cite{atlam2018blockchain}) are applied to the traditional financial ecosystem.

To give a broader understanding of how DeFi products are currently used in practice, we categorize DeFi products into lending services, assets, decentralized exchanges, derivatives and payment networks. For each of these categories, we provide examples and give an overview of how much funds are currently invested in products of this kind. Additionally, we provide an overview of current governance and economic topics in the field in Section~\ref{sec:economics}.

Lastly, we investigate the main challenges and possible solutions for DeFi products in a comprehensive way in Section~\ref{sec:challenges}. Here, we focus on the importance of secure smart contract implementation and analyze the consequences of bugs in the code or economic design of DeFi products. Additionally, we take a closer look at how the limited scalability of the Ethereum blockchain impacts the smart contract based applications and what countermeasures can be applied. Other challenges that we discuss are the secure usage of oracles, challenges concerned with the regulation of DeFi and how to onboard users easily.

\subsection{Related Work}
\label{subsec:related}
Concurrently to our work, Harvey et al.~\cite{HarveyRS20} surveyed financial infrastructures including DeFi.
While we analyze the DeFi ecosystem as a whole, another line of work focusses on single aspects.
Moin et al.~\cite{MoinSS20}, Klages{-}Mundt et al.~\cite{Klages-MundtHGL20} and Clark et al.~\cite{ClarkDM20} study the fundamentals of stablecoins which play an important role in DeFi protocols.
Pernice et al.~\cite{PerniceHPFE019} analyzes the stabilization of cryptocurrencies and systematically surveys existing DeFi products.
Daian et al.~\cite{DaianGKLZBBJ20} as well as Qin et al.~\cite{QinZLG20} study the security aspect of DeFi and show real-world attacks.

\section{Potential of DeFi}
\label{sec:potential}

Decentralized finance, rather than a single service, represents a whole ecosystem of financial services realized through smart contracts deployed on public distributed ledgers. This new approach aims to decentralize the current financial system by providing services and applications that are characterized by a detachment from trusted intermediaries, hence, enabling trustless, peer-to-peer transactions~\cite{ammous2015economics}. Instead of relying on traditional providers of financial services, which are accompanied by high costs, lengthy processes and a lack of transparency, DeFi realizes decentralized financial services. This way, applications such as lending, derivatives and trading are automated and executed trustlessly, transparently and securely~\cite{holotiuk2017impact}. Through the employment of publicly available  open protocols, decentralized apps (dApps) and smart contracts, DeFi enables individuals to play both sides of financial transactions, consuming and providing services, thus, “doing to finance what the internet did to the media by democratizing the creation of new financial instruments”~\cite{janashia2019}.

\subsection{Advantages of the DeFi Ecosystem}
\label{subsec:advantages}
The main advantages of DeFi services in comparison to traditional financial systems are that they are (1) permissionless, (2) trustless, (3) transparent, (4) interconnected, (5) decentrally governed and (6) they enable self-sovereignty for their users. 

\textbf{Permissionless:} Public blockchains are designed to be open, meaning that they do not specify access rules and anyone can interact with it~\cite{shoeb2019}. DeFi applications built on public blockchains inherit these properties by default. Permissions can be added only if the DeFi platforms are based on private blockchains with additional access restrictions or by specifying additional permissions in the code \cite{grech2017blockchain}. This also opens up the possibility for everyone to engage in speculation or margin trading, which is not available for everyone when using centralized platforms, as certain restrictions apply~\cite{xu2017enabling}.

\textbf{Trustless:} While distributed ledgers do not rely on a single operator as a trust agent, they distribute trust across a network of nodes instead \cite{anjum2017}. The security of the system is based on the assumption that enough nodes in the network behave honestly such that they can reach consensus on the validity of the recorded transactions \cite{chen2019decentralized}. Consensus guarantees that the system is immutable (meaning that transactions cannot be changed, added or deleted in retrospect) and censorship resistant (sometimes referred to as liveness), which guarantees that new valid transactions will be included eventually. 

\textbf{Transparent:} Most public distributed ledgers provide transparency by default, since all transactions stored in the blockchain are publicly visible. Transaction senders and receivers are identified with pseudonyms while transaction values and transmitted data are sent in the clear. Unless additional privacy measures are taken, it has been shown that transactions can be linked and users can also be identified~\cite{meiklejohn2013fistful}. For DeFi applications, the underlying transparency means that all usage and the stored funds are public at any point in time.    

\textbf{Interconnected:} Some blockchain ecosystems like Ethereum provide powerful programming tooling that is utilized for DeFi services. Complex applications, including auctions, voting and trading, can be built with smart contracts. Their features can be called by users and smart contracts, making it possible to easily connect, stack or combine existing applications without additional programming efforts. Combining contracts to create new types of services is often regarded as the \emph{Lego} property of DeFi protocols. This homage originates from using smart contracts as building blocks, combining them to create  more functional constructs. 

\textbf{Decentrally governed:} Not restricted to the DeFi space but highly prevalent in it is the aspect of turning smart contracts into decentralized autonomous organizations (DAOs). By enabling the community to suggest legislation and vote based on their stake in the project, governance is distributed. The three main aspects of governance, incentive compatibility, accountability and transparency, are realized by making users and investors responsible for the well-being of the ecosystem.

\textbf{Enabling self-sovereignty:} As no central authority controls and organizes access to the decentralized environment, users themselves manage their personal data and custody of their funds. As users can store their own access tokens, i.e., signing keys and authorize their own transactions, they are fully self-reliant. While this, on the one hand, prevents central services from stealing or confiscating funds, self-reliant users, on the other hand, are not protected against losing their access tokens since only they have the ability to restore them.

\begin{figure*}
	\centering
	\includegraphics[width=\textwidth]{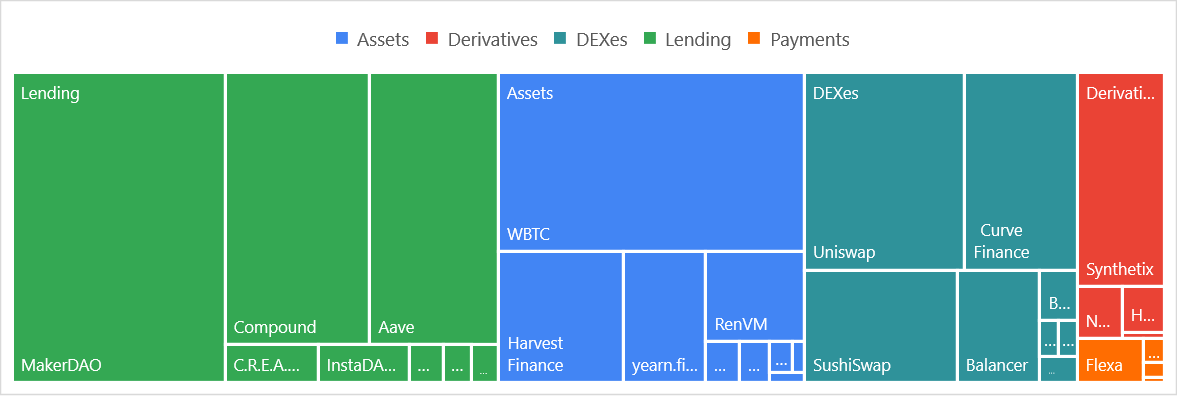}
	\caption{Value stored in financial DeFi Products based on data from \url{https://defipulse.com/}}
	\label{fig:treemap}
\end{figure*}

\subsection{Overview of Financial Services}
\label{subsec:categories}
In this section, we explain the potential of DeFi applications by presenting exemplary use cases. We classify the existing approaches into five categories: lending platforms, asset handling platforms, decentralized exchanges (DEXes), derivative services and payment networks. \figurename~\ref{fig:treemap} shows the market value of services in each category\footnote{Unless otherwise specified, all numbers are taken from \url{https://defipulse.com/} as of 12/01/2020.}. While more categories are emerging, we focus on these most important five categories.

\textbf{Lending Platforms:} 
\label{category:lending}
	Decentralized lending services make up the largest class of DeFi products with a total number of \$6.3B locked funds. They offer loans to businesses or individuals using smart contracts as intermediaries and negotiators. By doing so, the “regular” intermediaries required in centralized lending are eliminated. These smart contracts automate the lending and borrowing agreement, including rules for setting interest rates. The largest examples in this category are Maker~\cite{Maker} (\$2.75B), Compound~\cite{Compound} (\$1.6B) and Aave~\cite{Aave} (\$1.46B).
	
	One interesting mechanism that has been introduced in DeFi is the concept of \emph{flash loans}.
	It describes how a loan of any size can be received, used and payed back in a single transaction. No collateral is required for this type of loan. Should the debtor fail to pay back the loan, the transaction will be reverted, and the sender will only have to pay for transaction fees. Undesirable trades using this mechanism enabled numerous infamous attacks on DeFi protocols. 

\textbf{Assets:} 
\label{category:assets}
	Traditional assets mainly serve as capital backing for companies. Also, virtual assets within the cryptocurrency space have similar purposes. However, as they are created, stored and traded over the blockchain, they are inherently public, and their movements are transparent. Using smart contracts, they can be traded automatically and censorship-free. Assets make up the second largest Defi class in terms of locked value. Popular assets in the DeFi space are WBTC (wrapped Bitcoin) with a market value of \$2.28B and yToken by Yearn.finance~\cite{YearnFinance} (\$449M).  

\textbf{Decentralized Exchanges (DEXes):} 
\label{category:dex}
	Services that focus on decentralized cryptocurrency and token exchange are often classified as DEXes and make out the third largest class of DeFi products with around \$3.5B locked funds. DEXes work similar to a stock exchange, but instead of being run by a central provider, the exchange is operated by a smart contract deployed on blockchains like Ethereum. The lack of a centralized authority means the rules and regulations for trading are predetermined in the smart contract code and users have to interact with the contract in order to trade assets. Usually, the contract additionally handles the user funds during the trading process to ensure the correct payout.
	
	Uniswap~\cite{Uniswap}, whose users locked up value worth over \$1.3B, is the largest example within this category. The Uniswap contract is publicly available on the blockchain any user can directly interact with it. The main mechanic behind Uniswap is liquidity pooling, which removes the need to handle order books. Users pay 0.3\% fees per trade, which are added to the used liquidity pool and increase revenue for liquidity providers. Another popular DEX service is Curve Finance~\cite{CurveFinance} with over \$900M currently locked capital, which focuses mainly on trading and lending of stablecoins.

\textbf{Derivate Services:} 
\label{category:derivates}
	DeFi derivatives build on smart contracts that derive value from the performance of an underlying entity such as bonds, currencies, or interest rates. Tokenized derivatives can be created without third parties and by-design prevent malicious influence. Popular examples for DeFi derivatives are Synthetix~\cite{Synthetix} (\$771.2M), Nexus Mutual~\cite{NexusMutual} (\$96M) and Erasure~\cite{Erasure} (\$4.9M).
	
\textbf{Payment Networks:}
\label{category:payments}
	Even simple financial tools like payments are decentralized to reduce the influence of central payment providers with the goal of creating an open finance ecosystem. Due to the nature of blockchain technology, users can exchange cryptocurrency securely and directly without the need for intermediaries. However, high fees and inflexible delays call for specialized services that optimize decentralized payments. The underlying technology for services like Lightning Network~\cite{Lightning} (\$19.9M) rely on payment channel technologies, xDAi~\cite{xDai} (\$4.9M) builds on sidechains instead. 

The above mentioned categories are not absolute and pose classification guidelines, as many aspects of DeFi are still subject to change. Moreover, many Defi services may be associated with more than one or even additional categories. Gnosis~\cite{Gnosis} for example, can be classified as both a DEX because of the Gnosis protocol and asset through conditional tokens (event-based assets).

In addition to the above mentioned categories, oracles expand the given view on DeFi offerings. Whenever real-world data such as a currency exchange rate or the outcome of an election is needed within a smart contract, oracle services such as ChainLink~\cite{Chainlink} are used. ChainLink aims to provide tamper-proof data using a transparent network of audited oracle operators located in different geographic regions. The data provided by ChainLink includes price feeds for DeFi applications as well as randomness for gaming use cases and can be used for multiple blockchains. Inserting live data into a blockchain creates a number of challenges that are discussed in section \ref{subsec:oracles}. The data provided by oracles can be used for prediction markets like Augur~\cite{Augur}. Augur enables the global participation of betting on global events without middlemen as the funds and payouts are handled by smart contracts.  The emergence of a growing number of DeFi applications has created a variety of tooling that help to connect different DeFi services such as 0x~\cite{0x}. With 0x users can trade tokens on the Ethereum blockchain by querying multiple DEXes at once to get the best possible price.

Lastly, governance communities as provided by projects like Aragon~\cite{Aragon} experiment with decentrally governed communities that include majority voting for things like a manifesto or the adoption of new processes. Future phases of the Aragon project include the deployment of a court that is governed by a council in a transition phase and later by the community. In the following section, a more detailed view on decentralized governance as well as the economics behind it is given.

\section{Economics and Governance}
\label{sec:economics}

\subsection{Decentralized Governance}
\label{subsec:governance}
Decentralized governance models in the DeFi space are prominent. Major protocols and exchanges use them to update their policy regularly, allowing stakeholders to vote. The voting power depends on the number of governance tokens a stakeholder owns. A recent example is Uniswap. The airdrop (i.e., transfer of tokens to eligible wallets) by the largest DEX benefited users who used the service prior to a defined point in time. Any owner of the token is eligible to vote with their wallet, using the Uniswap interface. Not only exchanges use this model, but also the lending platform Maker has been using this approach for a while now. The difference here is that the Maker governance token (MKR) is deflationary due to burning dynamics, while the Uniswap governance token (UNI) has inflationary properties.

These various approaches lead to different effects: while a UNI holder loses voting power if they are not actively mining the tokens by providing liquidity, MKR holders tend to become more influential over time without active interaction with the protocol. The first approach benefits those who might have come late but are active. The latter approach primarily benefits legacy owners. Both inflationary and deflationary dynamics make sense for smart contracts as they serve their respective use case: MKR tokens need to be burned in order to stabilize DAI, and Uniswap aims to gradually become more decentralized by allowing users to mint. These statements have to be carefully reflected in light of the respective token prices as they are not stable. Like other crypto assets, these governance tokens are traded.

Currently, the voting power distribution does not (yet) support the assumption of a higher degree of centralization in the deflationary governance token system. This assumption is grounded on the idea that in a deflationary context, a single token becomes more valuable over time and that there are individuals who anticipate this by only holding the tokens. The novelty of the UNI token explains the currently higher concentration of voting power in Uniswap. An analysis of data provided by etherscan\footnote{Data obtained from etherscan.io on the 1\textsuperscript{st} of December 2020.} is used to determine and compare the concentration of voting power. The top 100 wallets which cannot be assigned to a major exchange, investment fund or a smart contract currently hold 26.4\% of MKR tokens. As for UNI, these users hold 42.7\%, which translates to a higher concentration of voting power. Including the smart contracts and exchanges, the numbers increase to 87.3\% for MKR and 93.7\% for UNI. However, these numbers include possibly thousands of holders who exclusively own their respective assets on exchanges. A distinction between these two perspectives is important because tokens locked up in exchanges will often not be used to vote.

Although DeFi products are built on top of a decentralized network, and decentralized governance mechanisms are in place, we identified a weakness based on the high dependence on the Tether stablecoin (USDT).
Because USDT is the primary stablecoin in use, it is a de facto central gateway. This introduction of counterparty risk in the decentralized financial systems can have consequences: accessibility for many decreases and the offramping becomes significantly harder, which might cause the ecosystem to halt its growth and possibly revert it until a suitable replacement is found. This dependence has to be considered because Tether already had issues with liquidity, asset security and alleged price manipulation in the past.

\subsection{Economics}
\label{subsec:economics}
\begin{figure}
 	\centering
	\includegraphics[width=\columnwidth]{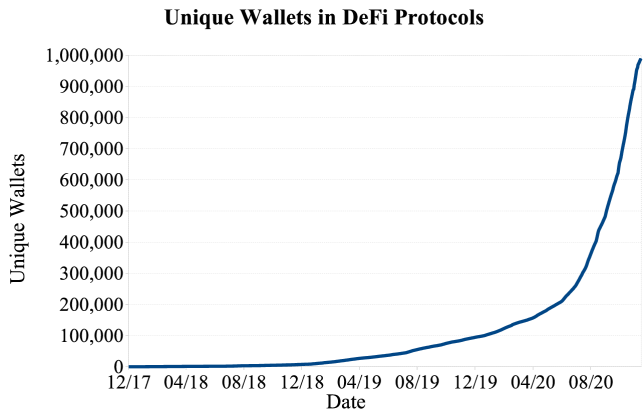}
	\caption{Daily growth of unique wallets in UniSwap, Compound, Kyber and 24 additional DeFi protocols from December 2018 to today. Based on data from Dune Analytics~\cite{DuneAnalytics}.}
	\label{fig:walletGrowth}
\end{figure}

The DeFi space has seen rapid growth in 2020, and current trends in Q3 2020 are described as the  largest bull run since the ICO boom in 2017~\cite{consensys2020q3}. In the time from November 2018 to November 2020, the locked value in DeFi protocols has increased close to 80 times from \$175M to \$13.9B. This growth in locked up value (collaterals and lent DAI) stems not only from reinvested gains and increased Ether valuations but also due to a larger number of users. At the time of writing, almost one  million\footnote{This number is an overestimation as users can create multiple wallets.} DeFi users (measured in unique addresses) has been counted by Dune Analytics~\cite{DuneAnalytics}.
Although it is not a perfect metric, a growth in wallets of more than 1000\% in the past year shows a strong tendency. The development is visualized in \figurename~\ref{fig:walletGrowth}. 

\begin{figure}
	\centering
	\includegraphics[width=\columnwidth]{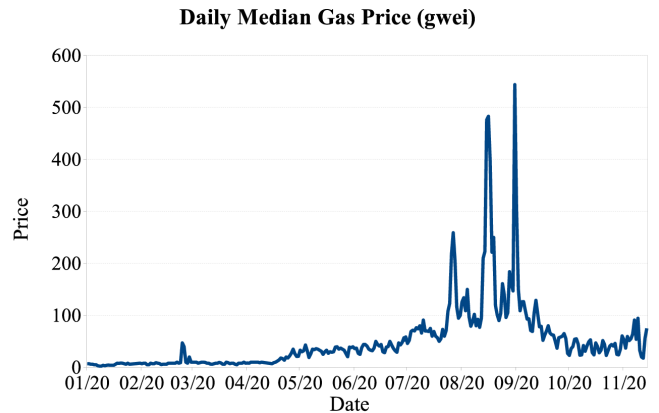}
	\caption{Daily median gas prices on the Ethereum network denominated in gwei. Based on data from Dune Analytics~\cite{DuneAnalytics}.}
	\label{fig:medianGasPrices}
\end{figure}

It is assumed that DeFi significantly influenced the transaction cost within the Ethereum network. Consequently, the incentive to avoid unnecessary transactions has increased: Gas usage of over 60 billion gas (daily) has been observed since June of 2020. With average gas prices between 40 and 540 gwei (1 gwei = $ 10^{-8}$\,ETH) in this period of time (see \figurename~\ref{fig:medianGasPrices}¸), the estimated gas burned in the second half of 2020 will exceed \$1.5M per day. This development is not sustainable in Ethereum (version 1), and its proof of work mechanism, as will be discussed in a later section. Especially automated trading protocols and DEX arbitrageurs in general, have caused a massive increase in the number of transactions in recent months. Prior to DEXes, arbitrage trading in the crypto sphere was more difficult. Provided with DeFi protocols, arbitrageurs can obtain financial gains for the service of evening out price discrepancies. 

While the metrics of this space are notable, they are shy of what conventional financial markets transact and consume on a daily basis. 
The DeFi space is still novel, unregulated and taxation for potential users remains unclear. Moreover, current protocols do not support any officially accepted ledger. These factors reduce the number of potential users joining the system. It can be expected that increased institutional and regulatory attention will be laid on decentralized finance, and consequently, the decentralization of the financial sector in the coming years. It is expected that a more regulated space will attract more users, as has been witnessed with the increasing adoption of cryptocurrencies.

\section{Challenges}
\label{sec:challenges}

This section discusses the most critical challenges that users and developers in the DeFi space face and provides an overview of potential solutions. 

\subsection{Security}
\label{subsec:security}
We identify three aspects of DeFi products that require special attention in terms of security: smart contract vulnerabilities, infrastructural risk and interdependence weaknesses. The past has shown that insufficient protection has resulted in massive financial loss. We point out some of the most prominent incidents.

First, DeFi products are built upon smart contracts dealing directly or indirectly with user funds. When more money is associated with a certain smart contract, it becomes more attractive for attackers. Thus, smart contracts can be seen as an equivalent to public bug bounty programs since every user that finds a bug inside the contract can exploit the vulnerability and potentially steal money. The fact that the contract code and all past interactions with it are transparently stored on the blockchain makes it even easier to find bugs. Therefore, smart contract developers must put much effort into programming contracts without vulnerabilities. Using well-known design patterns and best practices is a good starting point. Additional security audits from external parties may increase the trust in the correctness of a contract as well. Developers can also design a contract such that potential security patches might be applicable while the contract is running on the blockchain. However, such an update mechanism requires some form of governance, which effectively lowers the degree of decentralization. The past showed the massive impact of programming bugs in smart contracts, e.g., on the DAO and  parity wallets \cite{AnotherDayAnother2020,DeFiProjectOrigin2020,GarbageCryptoProduct,YFIFounderIncomplete2020}.

Second, the underlying infrastructure may have additional influences on the DeFi product, which needs to be considered when designing application-specific security mechanisms. For instance, the limited throughput of the Ethereum blockchain led to a congestion of the network in 2020 (see \figurename~\ref{fig:medianGasPrices}). Suppose a contract makes use of timeouts to ensure timely interaction by the participants. In that case, a congested network may result in users missing their timeouts since valid transactions from honest users might not be recorded in time~\cite{WinzerHF19}. Hence, the properties of the underlying consensus mechanism influence application-specific security properties.

Third, designing new protocols for the DeFi space requires special consideration. In particular, because of the opportunity of composing different DeFi products and creating new protocols based on existing ones. The security of a single protocol cannot be analyzed in a standalone model; influences of other protocols also need to be taken into account. We show this aspect by highlighting two specific attacks presented in prior work. The first attack, called frontrunning, was analyzed by Daian et al.~\cite{DaianGKLZBBJ20}. The term frontrunning comprises all scenarios where one party tries to get her transaction recorded before a competing transaction. Any attempt to frontrun may result in a so-called priority gas auction where users alternate in increasing their transactions' gas price to incentive miners to include their transactions first.

One example of frontrunning is arbitrage. By deliberately exploiting different exchange rates at DEXes, users can gain money via arbitrage.
Although arbitrage is also possible in traditional finance, using it in combination with smart contracts raises the treat to the next level. Batching several trading transactions into a single proxy contract allows the arbitrageur to execute the trades in an atomic way. This provides the advantage of executing either a successful arbitrage or paying only minimal fees for the unsuccessful contract execution.
Here, users benefit from the atomicity of smart contract execution similar to flash loans, as explained in Section~\ref{category:lending}.

It is tempting to think that frontrunning and priority gas auctions only affect application-layer security. Unfortunately, Daian et al.~\cite{DaianGKLZBBJ20} showed that application-layer attacks pose a threat to consensus-layer security as well. Rational consensus nodes might be incentivized to hold blocks back, fork the main chain or even rewrite history. Although for the latest attack, the miner must comprise enough mining power, depending on the values gained by rewriting, it might be profitable to do so.

Another attack vector described by Qin et al.~\cite{QinZLG20} is based on the usage of flash loans (cf. Section~\ref{category:lending}). However, flash loans cannot be considered as clearly benign or malicious. Instead, it depends on the borrower’s intention. Qin et al. explored use cases of flash loans as well as attacks exploited on the Ethereum blockchain~\cite{QinZLG20}. One attack named pump and arbitrage uses flash loans to manipulate exchange rates at one DEX in order to create an arbitrage opportunity.

A similar attack was based on the manipulation of DEX prices, which are used as price oracles by a lending service.
By trading for a high volume of sUSD token at two DEXes, another lending platform that uses these DEXes as price oracles act accordingly and decrease the exchange price for sUSD.
In the next step, the attacker converts ETH (Ethereum token) to sUSD for the lower price.
Finally, all sUSD are exchanged against ETH at a lending platform using the undistorted exchange rate.
As a result, the attacker possessed a lot more ETH than before this attack.
As already pinpointed in~\cite{QinZLG20}, flash loans do not open up new attacks in the DeFi space but rather amplify these attacks since no collateral is required to execute them. Qin et al. name starting points for potential defenses, e.g., DEX might reject any trades based on flash loans, or a delay between different actions like price oracle requests and updates could be introduced.

Finally, due to the high interest in DeFi, the system becomes more and more attractive to attacks and scams. In September 2020~\cite{bitcoinNews}, it was revealed that the FEW token was orchestrated as a pump and dump scheme. Influencers distributed information about this token to trigger other users to invest in the token. Once the token price was high enough, the core investors sold their shares to gain a high return on investment. 

\subsection{Limited Scalability}
\label{subsec:scalability}
Blockchain technology and its applications suffer from a limited transaction throughput, which is often viewed as the main hurdle for mass adoption of this technology~\cite{buterinInterview}. The underlying reason is that blocks in the ledger only have limited space shared by transactions, smart contract deployments and contract function invocations. Hence, when many applications and their users compete for the limited space in blocks, miners pick the transactions that offer the highest fees. Thus, users have to accept either very high fees or long confirmation delays for new transactions. 

Ethereum is the primary choice for DeFi applications because of its programmability, extensive community and a large variety of developer tooling. However, because of its limited scalability, Ethereum cannot handle the growing number of users and emerging DeFi applications. A study conducted by the German digital industry association Bitkom states that it is highly questionable whether Ethereum is a viable platform for DeFi, especially when even more users enter the system~\cite{grigoDecentralizedFinanceDeFi}. When the demand for DeFi applications was spiking in late summer 2020 (see \figurename~\ref{fig:medianGasPrices}), the Ethereum transaction fees increased dramatically and made many other applications on the Ethereum blockchain impossible.

This phenomenon can impact tokens like the Basic Attention Token, which is rewarded to those watching ads. Owners can tip content creators or service providers they use~\cite{BAT}. This model works more effectively if Ethereum transaction costs are negligible compared to the volume of the transaction. As a consequence, with increasing gas costs, the utility of the BAT decreases. However, this is impractical for applications that only control small monetary amounts and impacts the security of many existing dApps on the Ethereum blockchain. One reason is that many smart contracts use timelocks, i.e., to specify deadlines for users to react or complain. These contracts assume that timeouts are sufficiently long for honest users to react for reasonable fees - which might be prevented when the blockchain becomes congested.

The challenge of limited scalability is tackled on two different layers. Layer-1 solutions aim for improving the consensus mechanism of blockchain technologies~\cite{BachMZ18b}. The EOS blockchain~\cite{EOS}, for example, uses an alternative consensus mechanism called delegated proof of stake (DPoS) to achieve more scalability.
Other approaches include applying sharding techniques already used in the context of database management~\cite{obasanjo2009}. Sharding splits the state of the whole blockchain into several independent units called shards. Instead of validating each new transaction by all nodes of the consensus network, only those assigned to the specific shard are responsible for validating transactions belonging to this partition. This enables the network to process transactions of different shards in parallel and thus increases the overall throughput. Sharding is incorporated in other blockchain technologies like Zilliqa~\cite{zilliqa} and Elrond~\cite{elrond}. The introduction of sharding techniques is also planned for the Ethereum blockchain as part of four phases towards Ethereum 2.0~\cite{EthereumBeaconChain2020}. Current developments, however, state that the native sharded execution of transactions is de-emphasized for the time being.

In contrast to Layer-1 solutions, Layer-2 techniques tackle the application layer's scalability issue without requiring any changes to the underlying consensus mechanism. This provides the opportunity to add these solutions to already existing blockchain technologies without modifying the backbone system.
Current Layer-2 solutions for Ethereum increase the transaction throughput from currently 15 transactions per second (tps) to up to 2,000 tps, lower transaction fees and improve latency. Layer-2 scaling solutions applicable to DeFi use cases require the execution of arbitrary smart contracts, which makes current scaling solutions such as payment channels \cite{Lightning,RaidenNetwork} and Plasma \cite{PlasmaGroup} much less relevant. Solutions tailored towards the challenges of DeFi use cases include zkRollups, optimistic rollups and optimized optimistic rollups \cite{NewWayScale,LoopringZkRollupExchange,starkwareSTARKsMainnet2020}. Further, research approaches such as CommitTEE are leveraging secure enclaves to combine benefits such as very high throughput and zero gas fees.  Trusted execution devices eliminate disadvantages such as high computational overhead and regular on-chain checkpoints of the history of the off-chain transactions \cite{erwigCommiTEEEfficientSecure2020}. \snote{Shouldnt we not also mention here some other TEE-Based solutions? For example the Bitcontracts or so?}

DeFi projects are already actively using Layer-2 scaling solutions.  Rollups, in particular, are already actively used or in Beta testing in different DeFi projects, including StarkEx (zkRollups), Loopring (zkRollups), IDEX (optimistic rollups), DeversiFi (optimized optimistic rollups) and Unipig (optimistic rollups)~\cite{STARKEx,LoopringZkRollupExchange,idexIDEXHighPerformanceDecentralized,subscribeDeversiFiDecentralizedEthereum,UnipigExchange}. 
It can therefore be assumed that rollups are the technology of choice for current DeFi projects.
Grigo et al.~\cite{grigoDecentralizedFinanceDeFi} state that the transition to Ethereum 2.0 is essential to handle the growing user demand.
Since the native sharded execution of transactions is de-emphasized for the time being, the integration of rollups is planned to scale Ethereum in the near future.

\subsection{Oracles}
\label{subsec:oracles}
While the interaction between two on-chain entities like smart contracts is simple, transferring information from external sources like websites to a smart contract creates new challenges. Many DeFi products rely on external information like exchange rates, which is provided by so called oracles. Since the data originating from these oracles impact the behavior of smart contracts and users, the challenges posed by transferring external data on-chain is a major concern. In particular, the security of these DeFi products is based on the reliability, accuracy and correctness of the provided information from oracles. Therefore, oracles are evaluated based on their transparency, accountability and the required level of trust.
Popular oracle-based DeFi products are Maker, Compound, AmpleForth~\cite{Ampleforth} and Synthetix~\cite{Synthetix}. ChainLink~\cite{Chainlink} is even providing an network of oracles making information accessible via its API. Its growing relevance is likely connected to significant partnerships, e.g., with Google, Oracle or Salesforce~\cite{Chainlink-ecosys}.

We elaborate on the usage of oracles by describing how the Maker project addresses some of the challenges above by combining inputs from multiple sources instead of relying on data from a single source. For each type of external data, a whitelisted set of oracles is determined that frequently provides samples. These samples are consolidated by an aggregator, which produces the final data sent to the platform. Hence, the reliability of oracle data is improved by replication. The used oracles should be independent such that a malfunction of a single oracle does not influence other oracles. The aggregator computes the median of the reported samples in order to cancel out any large deviation.
Furthermore, Maker provides the possibility to update the whitelist of oracles resulting in exchanging oracles. The update is based on the decision of the governors controlling the MKR tokens.

Liu and Szalachowski~\cite{LiuSza20} conducted a study of DeFi oracles presenting large-scale measurements about price volatility, price deviation, failures and transaction activity analysis. Moreover, the authors propose recommendations for the design of oracle solutions. First, every oracle should clearly state information like data sources, update frequency, and price deviation description. Such general information can easily be provided and massively increases the transparency as it allows users to understand where deviations of different sources originate. 
Second, since oracles act as trusted third parties, it should be possible to hold them accountable for misbehavior like missing reports or high price deviation. This can be achieved by incorporating incentives into the design of an oracle solution. For example, oracle operators gain rewards when the frequency and accuracy of their feeds are adequate. Punishment for poor performance can be realized by forgoing rewards and possibly even withdrawal of crypto assets from a priorly provided pool.

\subsection{Regulation}
\label{subsec:regulationChallenges}
Creating global uniform standards for the regulation of crypto-economics could alleviate risks like censorship or collusion but are effectively nonexistent as of today. Most existing regulatory concepts are yet primarily concerned with the classification of tokens for taxation purposes. Liechtenstein and United States authorities act as global role models in doing so, and overall, regulators have increased clarity by following their example~\cite{nystrom2019}. For DeFi, it is yet unclear how generated income is regulated. The legal status of the entire ecosystem as such is not clearly defined. Questions about the potential for abuse or illicit usages arise. However, it is often unclear if the ecosystem can even face a shutdown. Penalizing certain usage is hard due to the aspects of self-reliance and decentral finalization of transactions. There is a significant gap between governance and external regulation to fill concerning DeFi on Ethereum. Moreover, the lack of know your customer (KYC) processes in DeFi ecosystems makes it harder for regulators to accept it as an official financial space. KYC practices can barely be enforced. As a consequence, regulators are confronted with the great challenge of not inhibiting innovation too strongly when regulating DeFi. \cite{yeung2019regulation} states that a balance between legal and technical code sustains interactions of different dimensions (economic, political, social) without harming the community.

In September of  2020, the European Commission presented a draft for the regulation of ``crypto assets'' (digital, blockchain-based assets), which is expected to be in force by 2023. The regulation ``Markets in Crypto-assets'' (``MiCA''), which is directly applicable for all European member states, describes the most extensive regulation of digital assets to date. As for DeFi, it is not yet clear which consequences this draft brings. While the proposal covers most types of crypto assets and categorizes them differently, DeFi tokens are not explicitly dealt with. The DAI stablecoin can be classified as a so-called asset-referenced Token according to the draft~\cite{MiCA}. The classification is justified by the soft peg to the US Dollar. Still, lots of tokens and contracts can be considered ``issuer-less''. This is a key issue in this context. It is likely that smart contracts in the DeFi space can be classified as crypto asset service providers at some point. However, conclusive legal research has to be done in order to clarify the relationship to DeFi further. 

\subsection{On- And Offramping}
\label{subsec:on/offramping}
Usability and user experience can determine the fate of projects. As DeFi was initially designed from crypto-natives, it was also designed for them. By now, the design of several dApps has significantly improved. However, the terms of use are frequently explained on a high technical level or deeply embedded in financial jargon. In the longterm, this poses a threat to the mass adoption of several DeFi projects. A possible solution can be to guide inexperienced users through the workflow in a more tutoring way, displaying implications in the process (while providing a link to an explanation).

On- and offramps refer to the methods to exchange traditional assets for crypto-assets and vice versa. Centralized exchanges are based on trust in an intermediary, require authentication via KYC practices, have limited scalability, suffer from security issues, process transactions off-chain and charge significant fees~\cite{Koenig}. Many of these shortcomings are equivalent to limitations that traditional banks face. The leading centralized exchanges are Coinbase, Binance and Kraken. To enable seamless on- and offramps, these companies must evolve significantly to satisfy all customers' requirements.

\subsection{Privacy}
\label{subsec:privacy}
The fact that all data is public on the blockchain poses several challenges - ranging from ``transaction linkability, crypto-key management, issues with crypto-privacy resistance to quantum computing, on-chain data privacy, usability, interoperability, or compliance with privacy regulations, such as the GDPR''~\cite{BernabeCRMS19}. By default, all the transactions that take place on Ethereum are publicly visible. Although the system's addresses are pseudonyms, they are decodable using information from centralized exchanges about client identification and other metadata.

Since financial data is highly sensitive for many individuals, privacy is a relevant topic.
In particular, users aim for private transactions such that no unauthorized party can obtain information about users' financial activities.
Moreover, the decentralization, openness and integrity protection of blockchain technologies pose challenges for compliance with privacy regulations (i.e. the right to be forgotten).

However, several projects address these issues. Private transfers (as described in~\cite{Berg2019}) can be achieved using several techniques. For example, disconnecting the link between the sender and recipient of tokens is possible when using a mixer based on smart contracts~\cite{Tornadocash2019}. Rollups allow users to hide smart contracts~\cite{ETHOpt} and Ernst \& Young shared an open-source repository called Nightfall that uses zk-snarks to make Ethereum transactions private~\cite{EY2019}. Still, the overall privacy supply on public blockchains has significant remaining challenges to cope with~\cite{BernabeCRMS19}. Bernabe et al.~\cite{BernabeCRMS19} argue that it is the users’ right to act anonymously in specific situations and that only by adhering to this right blockchains can provide a genuinely self-sovereign identity model.

\section{Summary}
\label{sec:summary}

DeFi as a whole will remain an interesting phenomenon and has lots of potential, growing continuously.
While early services focused on payments and trading solutions, the development shifted towards more advanced products realizing more complex financial services.
The Lego aspect intensifies this evolution even further.
Based on the increasing complexity, governance becomes more and more important.

The major challenges in the near future remain to be scalability and security.
In particular, the existing scaling issues pose the question whether or not Ethereum as the state-of-the-art DeFi platform is able to handle the growing demands.
Moreover, the regulatory uncertainties need to be solved.
A solution for KYC is not yet available and as a consequence, DeFi lacks proper recognition as a valuable financial service ecosystem in the public eye.

All in all, it can be expected that DeFi's growth can co-determine the growth of the blockchain sphere within the coming years as it motivates solutions and gives individuals the opportunity to access services when unbanked.

\section{Acknowledgments}

We want to thank Leon Erichsen whose research work guided us to gain a productive perspective.
This work was partly funded by the iBlockchain project (grant nr. 16KIS0902) funded by the German Federal Ministry of Education and Research (BMBF), by the Deutsche Forschungsgemeinschaft (DFG, German Research Foundation) – SFB 1119 – 236615297 (Project S7), and by the German Federal Ministry of Education and Research and the Hessian Ministry of Higher Education, Research, Science and the Arts within their joint support of the National Research Center for Applied Cybersecurity ATHENE.

\bibliographystyle{abbrv}
\bibliography{bibliography}

\begin{thebibliography}{10}

\bibitem{0x}
0x.
\newblock \url{https://0x.org/}.
\newblock \footnote{(Accessed on 12/17/2020)\label{fn:accessed}}.

\bibitem{Aave}
Aave – open source defi protocol.
\newblock \url{https://aave.com/}.
\newblock \footref{fn:accessed}.

\bibitem{Ampleforth}
Ampleforth.
\newblock \url{https://www.ampleforth.org/}.
\newblock \footref{fn:accessed}.

\bibitem{AnotherDayAnother2020}
Another day, another hack.
\newblock
  \url{https://cryptoslate.com/another-day-another-hack-2m-in-dai-drained-from-ethereum-defi-app-akropolis/}.
\newblock \footref{fn:accessed}.

\bibitem{Augur}
Augur.
\newblock \url{https://augur.net/}.
\newblock \footref{fn:accessed}.

\bibitem{BAT}
Basic attention token.
\newblock \url{https://basicattentiontoken.org/}.
\newblock \footref{fn:accessed}.

\bibitem{Chainlink}
Chainlink.
\newblock \url{https://chain.link/}.
\newblock \footref{fn:accessed}.

\bibitem{Chainlink-ecosys}
Chainlink ecosystem.
\newblock \url{https://chainlinkecosystem.com/ecosystem/}.
\newblock \footref{fn:accessed}.

\bibitem{Compound}
Compound.
\newblock \url{https://compound.finance/}.
\newblock \footref{fn:accessed}.

\bibitem{buterinInterview}
Crypto bites: Chat with ethereum founder vitalik buterin - youtube.
\newblock \url{https://www.youtube.com/watch?v=u-i_mTwL-FI&feature=youtu.be}.
\newblock \footref{fn:accessed}.

\bibitem{CurveFinance}
Curve.fi.
\newblock \url{https://www.curve.fi/}.
\newblock \footref{fn:accessed}.

\bibitem{DeFiProjectOrigin2020}
Defi project origin protocol exploited for \$7.7 million.
\newblock
  \url{https://coingeek.com/defi-project-origin-protocol-exploited-for-7-7-million/}.
\newblock \footref{fn:accessed}.

\bibitem{DuneAnalytics}
Dune analytics.
\newblock \url{https://duneanalytics.com/}.
\newblock \footref{fn:accessed}.

\bibitem{EOS}
Eosio - blockchain software architecture.
\newblock \url{https://eos.io/}.
\newblock \footref{fn:accessed}.

\bibitem{Erasure}
Erasure.
\newblock \url{https://erasure.world/}.
\newblock \footref{fn:accessed}.

\bibitem{EthereumBeaconChain2020}
Ethereum 2.0 beacon chain goes live as 'world computer' begins long-awaited
  overhaul.
\newblock
  \url{https://www.coindesk.com/ethereum-2-0-beacon-chain-goes-live-as-world-computer-begins-long-awaited-overhaul}.
\newblock \footref{fn:accessed}.

\bibitem{GarbageCryptoProduct}
Garbage crypto product dies immediately after launch.
\newblock
  \url{https://gizmodo.com/garbage-crypto-product-dies-immediately-after-launch-1844718822}.
\newblock \footref{fn:accessed}.

\bibitem{Gnosis}
Gnosis.
\newblock \url{https://gnosis.io/}.
\newblock \footref{fn:accessed}.

\bibitem{idexIDEXHighPerformanceDecentralized}
{IDEX}.
\newblock \url{https://idex.io}.
\newblock \footref{fn:accessed}.

\bibitem{Lightning}
Lightning network.
\newblock \url{https://lightning.network/}.
\newblock \footref{fn:accessed}.

\bibitem{LoopringZkRollupExchange}
Loopring.
\newblock \url{https://loopring.io/}.
\newblock \footref{fn:accessed}.

\bibitem{Maker}
Maker.
\newblock \url{https://makerdao.com/}.
\newblock \footref{fn:accessed}.

\bibitem{NewWayScale}
A new way to scale - optimized optimistic rollup.
\newblock \url{https://blog.idex.io/all-posts/o2-rollup-overview}.
\newblock \footref{fn:accessed}.

\bibitem{Aragon}
Next-level communities run on aragon.
\newblock \url{https://aragon.org/}.
\newblock \footref{fn:accessed}.

\bibitem{NexusMutual}
Nexus mutual.
\newblock \url{https://nexusmutual.io/}.
\newblock \footref{fn:accessed}.

\bibitem{PlasmaGroup}
Plasma group.
\newblock \url{https://plasma.group/}.
\newblock \footref{fn:accessed}.

\bibitem{RaidenNetwork}
Raiden {Network}.
\newblock \url{https://raiden.network/}.
\newblock \footref{fn:accessed}.

\bibitem{STARKEx}
{STARKEx}.
\newblock \url{https://starkware.co/product/starkex/}.
\newblock \footref{fn:accessed}.

\bibitem{Synthetix}
Synthetix.
\newblock \url{https://www.synthetix.io/}.
\newblock \footref{fn:accessed}.

\bibitem{UnipigExchange}
Unipig {Exchange}.
\newblock \url{https://unipig.exchange/welcome}.
\newblock \footref{fn:accessed}.

\bibitem{Uniswap}
Uniswap.
\newblock \url{https://uniswap.org/}.
\newblock \footref{fn:accessed}.

\bibitem{xDai}
Welcome to {xDai} stake - {xDai} stake.
\newblock \url{https://www.xdaichain.com/}.
\newblock \footref{fn:accessed}.

\bibitem{YearnFinance}
yearn.
\newblock \url{https://yearn.finance/}.
\newblock \footref{fn:accessed}.

\bibitem{YFIFounderIncomplete2020}
Yfi founder's incomplete defi protocol eminence exploited, attacker drained
  \$15m and then returned \$8m.
\newblock
  \url{https://www.theblockcrypto.com/post/79061/yfi-eminence-defi-protocol-exploited}.
\newblock \footref{fn:accessed}.

\bibitem{zilliqa}
The zilliqa technical whitepaper.
\newblock \url{https://docs.zilliqa.com/whitepaper.pdf}, August 2017.

\bibitem{elrond}
Elrond.
\newblock \url{https://elrond.com/assets/files/elrond-whitepaper.pdf}, June
  2019.

\bibitem{ammous2015economics}
S.~Ammous.
\newblock Economics beyond financial intermediation: Digital currencies’
  potential for growth, poverty alleviation and international development.
\newblock \url{https://ssrn.com/abstract=2832738}.

\bibitem{anjum2017}
A.~{Anjum}, M.~{Sporny}, and A.~{Sill}.
\newblock Blockchain standards for compliance and trust.
\newblock {\em IEEE Cloud Computing}, 4(4):84--90, 2017.

\bibitem{atlam2018blockchain}
H.~F. Atlam, A.~Alenezi, M.~O. Alassafi, and G.~Wills.
\newblock Blockchain with internet of things: Benefits, challenges, and future
  directions.
\newblock {\em International Journal of Intelligent Systems and Applications},
  10(6):40--48, 2018.

\bibitem{BachMZ18b}
L.~M. Bach, B.~Mihaljevic, and M.~Zagar.
\newblock Comparative analysis of blockchain consensus algorithms.
\newblock In {\em International Convention on Information and Communication
  Technology, Electronics and Microelectronics}, pages 1545--1550, 2018.

\bibitem{BernabeCRMS19}
J.~B. Bernab{\'{e}}, J.~L. C{\'{a}}novas, J.~L.~H. Ramos, R.~T. Moreno, and
  A.~F. Skarmeta.
\newblock Privacy-preserving solutions for blockchain: Review and challenges.
\newblock {\em {IEEE} Access}, 7:164908--164940, 2019.

\bibitem{ethereum}
V.~Buterin et~al.
\newblock Ethereum whitepaper.
\newblock \url{https://ethereum.org/en/whitepaper/}.
\newblock \footref{fn:accessed}.

\bibitem{Tornadocash2019}
T.~Cash.
\newblock Introducing private transactions on ethereum now!
\newblock
  \url{https://medium.com/@tornado.cash/introducing-privatetransactions-on-ethereum-now-42ee915babe0},
  August 2019.
\newblock \footref{fn:accessed}.

\bibitem{chen2019decentralized}
Y.~Chen and C.~Bellavitis.
\newblock Decentralized finance: Blockchain technology and the quest for an
  open financial system.
\newblock {\em Stevens Institute of Technology School of Business Research
  Paper}, 2019.

\bibitem{ClarkDM20}
J.~Clark, D.~Demirag, and S.~Moosavi.
\newblock Demystifying stablecoins.
\newblock {\em {ACM} Queue}, 18(1):39--60, 2020.

\bibitem{MiCA}
E.~Commission.
\newblock Proposal for a regulation of the european parliament and of the
  council on markets in crypto-assets (mica), November 2020.

\bibitem{DaianGKLZBBJ20}
P.~Daian, S.~Goldfeder, T.~Kell, Y.~Li, X.~Zhao, I.~Bentov, L.~Breidenbach, and
  A.~Juels.
\newblock Flash boys 2.0: Frontrunning in decentralized exchanges, miner
  extractable value, and consensus instability.
\newblock In {\em {IEEE} Symposium on Security and Privacy}, pages 910--927.
  {IEEE}, 2020.

\bibitem{erwigCommiTEEEfficientSecure2020}
A.~Erwig, S.~Faust, S.~Riahi, and T.~Stöckert.
\newblock Commitee: An efficient and secure commit-chain protocol using tees.
\newblock Cryptology ePrint Archive, Report 2020/1486, 2020.
\newblock \url{https://eprint.iacr.org/2020/1486}.

\bibitem{grech2017blockchain}
A.~Grech and A.~F. Camilleri.
\newblock Blockchain in education, 2017.

\bibitem{grigoDecentralizedFinanceDeFi}
J.~Grigo, P.~Hansen, D.~A. Patz, and V.~Von~Wachter.
\newblock Decentralized {Finance} ({DeFi}) – {A} new {Fintech} {Revolution}?

\bibitem{HarveyRS20}
C.~R. Harvey, A.~Ramachandran, and J.~Santoro.
\newblock Defi and the future of finance.
\newblock \url{https://ssrn.com/abstract=3711777}.

\bibitem{holotiuk2017impact}
F.~Holotiuk, F.~Pisani, and J.~Moormann.
\newblock The impact of blockchain technology on business models in the
  payments industry.
\newblock 2017.

\bibitem{janashia2019}
N.~Janashia.
\newblock Introduction to decentralized finance aka ‘defi’.
\newblock
  \url{https://medium.com/@Nodar/introduction-to-decentralized-finance-aka-defiea4f12e6256d},
  2019.
\newblock \footref{fn:accessed}.

\bibitem{Klages-MundtHGL20}
A.~Klages{-}Mundt, D.~Harz, L.~Gudgeon, J.~Liu, and A.~Minca.
\newblock Stablecoins 2.0: Economic foundations and risk-based models.
\newblock In {\em Conference on Advances in Financial Technologies}, pages
  59--79. {ACM}, 2020.

\bibitem{Koenig}
F.~Koenig.
\newblock Crypto exchanges explained.
\newblock
  \url{https://medium.com/wysker/crypto-exchanges-explained-549b42b47832}.
\newblock \footref{fn:accessed}.

\bibitem{EY2019}
C.~Konda, M.~Connor, D.~Westland, Q.~Drouot, and P.~Brody.
\newblock Nightfall.

\bibitem{LiuSza20}
B.~Liu and P.~Szalachowski.
\newblock A first look into defi oracles.
\newblock {\em CoRR}, abs/2005.04377, 2020.

\bibitem{meiklejohn2013fistful}
S.~Meiklejohn, M.~Pomarole, G.~Jordan, K.~Levchenko, D.~McCoy, G.~M. Voelker,
  and S.~Savage.
\newblock A fistful of bitcoins: characterizing payments among men with no
  names.
\newblock In {\em Proceedings of the 2013 conference on Internet measurement
  conference}, pages 127--140, 2013.

\bibitem{MoinSS20}
A.~Moin, K.~Sekniqi, and E.~G. Sirer.
\newblock Sok: {A} classification framework for stablecoin designs.
\newblock In J.~Bonneau and N.~Heninger, editors, {\em Financial Cryptography
  and Data Security}, pages 174--197, 2020.

\bibitem{consensys2020q3}
E.~Muzzy, J.~Beck, and T.~Hay.
\newblock Defi report.

\bibitem{nakamoto}
S.~Nakamoto.
\newblock Bitcoin: A peer-to-peer electronic cash system.
\newblock Technical report, Manubot, 2019.

\bibitem{nystrom2019}
M.~Nystrom.
\newblock 2019 was the year of defi (and why 2020 will be too).
\newblock
  \url{https://consensys.net/blog/news/2019-was-the-year-of-defi-and-why-2020-will-be-too/}.
\newblock \footref{fn:accessed}.

\bibitem{obasanjo2009}
D.~Obasanjo.
\newblock Building scalable databases: Pros and cons of various database
  sharding schemes.
\newblock {\em Obsanjo’s technical blog}, 2009.

\bibitem{ETHOpt}
E.~Optimism.
\newblock Optimism.
\newblock \url{https://medium.com/ethereum-optimism/optimism-cd9bea61a3ee}.
\newblock \footref{fn:accessed}.

\bibitem{PerniceHPFE019}
I.~G.~A. Pernice, S.~A. Henningsen, R.~Proskalovich, M.~Florian, H.~Elendner,
  and B.~Scheuermann.
\newblock Monetary stabilization in cryptocurrencies - design approaches and
  open questions.
\newblock In {\em Crypto Valley Conference on Blockchain Technology}, pages
  47--59. {IEEE}, 2019.

\bibitem{QinZLG20}
K.~Qin, L.~Zhou, B.~Livshits, and A.~Gervais.
\newblock Attacking the defi ecosystem with flash loans for fun and profit.
\newblock {\em CoRR}, abs/2003.03810, 2020.

\bibitem{shoeb2019}
S.~Shoeb.
\newblock Decentralization disrupting the finance ecosystem.
\newblock
  \url{https://medium.com/datadriveninvestor/compound-vs-nuo-vs-dharma-vs-maker-whichone-is-the-best-d85d5d614bb1}.
\newblock \footref{fn:accessed}.

\bibitem{starkwareSTARKsMainnet2020}
StarkWare.
\newblock {STARKs} over {Mainnet}.
\newblock \url{https://medium.com/starkware/starks-over-mainnet-b83e63db04c0},
  June 2020.
\newblock \footref{fn:accessed}.

\bibitem{subscribeDeversiFiDecentralizedEthereum}
{Subscribe}.
\newblock {DeversiFi} - {Decentralized} {Ethereum} {Exchange}.
\newblock \url{https://www.deversifi.com/}.
\newblock \footref{fn:accessed}.

\bibitem{swan2017anticipating}
M.~Swan.
\newblock Anticipating the economic benefits of blockchain.
\newblock {\em Technology innovation management review}, 7(10):6--13, 2017.

\bibitem{Berg2019}
D.~Z.~J. Williamson.
\newblock The aztec protocol.
\newblock \url{https://github.com/AztecProtocol/AZTEC/blob/master/AZTEC.pdf}.
\newblock \footref{fn:accessed}.

\bibitem{WinzerHF19}
F.~Winzer, B.~Herd, and S.~Faust.
\newblock Temporary censorship attacks in the presence of rational miners.
\newblock In {\em European Symposium on Security and Privacy Workshops}, pages
  357--366. {IEEE}, 2019.

\bibitem{xu2017enabling}
L.~Xu, N.~Shah, L.~Chen, N.~Diallo, Z.~Gao, Y.~Lu, and W.~Shi.
\newblock Enabling the sharing economy: Privacy respecting contract based on
  public blockchain.
\newblock In {\em ACM Workshop on Blockchain, Cryptocurrencies and Contracts},
  pages 15--21, 2017.

\bibitem{yeung2019regulation}
K.~Yeung.
\newblock Regulation by blockchain: the emerging battle for supremacy between
  the code of law and code as law.
\newblock {\em The Modern Law Review}, 82(2):207--239, 2019.

\bibitem{zile2018blockchain}
K.~Z{\=\i}le and R.~Strazdi{\c{n}}a.
\newblock Blockchain use cases and their feasibility.
\newblock {\em Applied Computer Systems}, 23(1):12--20, 2018.

\bibitem{bitcoinNews}
T.~Zimwara.
\newblock Defi token exposed as pump and dump scam in leaked telegram chat.
\newblock
  \url{https://news.bitcoin.com/defi-token-exposed-as-pump-and-dump-scam-in-leaked-telegram-chat/}.
\newblock \footref{fn:accessed}.

\end{thebibliography}

\end{document}